\documentclass[prd,10pt,aps,showpacs,natbib]{revtex4}

\usepackage[dvips]{graphicx}
\usepackage{textcomp}
\usepackage{amssymb}

\usepackage{graphicx,color}

\mathchardef\bigtilde="0365

\begin{document}

\title{Notes on the $\delta$-expansion approach to the 2D 
Ising susceptibility scaling}
\author{Hirofumi Yamada}\email{yamada.hirofumi@it-chiba.ac.jp}
\affiliation{%
Division of Mathematics and Science, Chiba Institute of Technology, 
\\Shibazono 2-1-1, Narashino, Chiba 275-0023, Japan}

\date{\today}

\begin{abstract}
{We study the scaling of the magnetic susceptibility in the square Ising model based upon the $\delta$-expansion in the high temperature phase.  The susceptibility $\chi$ is expressed in terms of the mass $M$ and expanded in powers of $1/M$.  The dilation around $M=0$ by the $\delta$ expansion and the parametric extension of the ratio of derivatives of $\chi$, $\chi^{(\ell+1)}/\chi^{(\ell)}$ is used as a test function for the estimation of the critical exponent $\gamma$ with no bias from information of the critical temperature.  Estimation is done with the help of the principle of minimum sensitivity and detailed analysis revealed that $\ell=0,1$ cases provide us accurate estimation results.  Critical exponent of the sub-leading scaling term is also estimated. 
}
\end{abstract}


\maketitle
\section{Introduction}
Recently, the $\delta$-expansion approach to the critical phenomena was studied within the high temperature expansion on the cubic Ising model \cite{yam}.   The work has concentrated on the behavior of the inverse temperature $\beta$ as a function of the mass parameter $M$, which is defined by the magnetic susceptibility $\chi$ and the second moment $\mu$.  The approach yielded results good enough for encouraging further exploration.   The present paper extends the approach to the magnetic susceptibility in the 2-dimensional (2D) square Ising model.  Within $\delta$-expansion approach applied to the high temperature series, we try to recover the critical behavior of the susceptibility.  Specifically, we will discuss unconventional way of estimating the exponent $\gamma$.  The new point resides in the 
examination of the ratio $\chi^{(\ell+1)}/\chi^{(\ell)}$ $(\ell=0,1,2,3,\cdots)$ where $\chi^{(\ell)}$ stands for the $\ell$th order derivatives with respect to $\log M^{-1}$,
\begin{equation}
\chi^{(\ell)}=\Big(x\frac{\partial }{\partial x}\Big)^{\ell}\chi,\quad x=1/M.
\end{equation}
The best choice of $\ell$, the order of the derivatives, will be determined in the high temperature expansion.  As will be explained in later (see (\ref{key1}) and (\ref{key2})), the ratio\begin{equation}
{\cal R}_{\ell}=\frac{\chi^{(\ell+1)}}{\chi^{(\ell)}}
\end{equation}
converges to $\gamma/(2\nu)$ in the critical limit $x\to \infty$ for $\ell$ of which value is not restricted at least formally.   In extracting the limiting value, we pay attention also to the best choice of $\ell$ for the estimation of $\gamma/(2\nu)$.

We like to remark the reader that, though $\chi$ is not obtained in a closed form in $\beta$, its mathematical structure has been studied and explored from long time ago in many prominent works (see the papers cited in \cite{mc,peli}).  The value of critical exponents are exactly known, for instance, as $\nu=1$ and $\gamma=7/4$.   The present work does not intend adding some new results along with the traditional direction.  Rather, our motivation is in the series expansion {\it in the mass} (but not in the temperature) and to develop a new approach based upon the $\delta$-expansion to the critical phenomena.  Unlike ordinary expansion in the temperature variable, the representation of thermodynamic quantities in the mass has an advantage that no knowledge on $\beta_{c}$ is required in the analysis of interested quantities.  We even consider the temperature as function of the mass.   Conventional approach takes thermodynamic quantities to be functions of $1-\beta/\beta_{c}$.  Then, to estimate critical quantities, one needs to know in the first place the precise value of $\beta_{c}$ because the accuracy of $\beta_{c}$ affects all estimation tasks.   In the present approach, no such bias is present.   As a typical example of application, we here employ the computation of the critical exponent $\gamma$.  We extract the exponent $\gamma/(2\nu)$ from the critical behavior of $\chi$  expressed in $1/M$ series.    We also stress that in the large mass expansion, the $\delta$-expansion method plays a crucial role.  By such a specific study, we expect a further development of the $\delta$-expansion method in the application of the critical phenomena and, hopefully, in other branches of physics.   

It is also well-known in the square Ising model that $\beta_{c}=\frac{1}{2}\log(1+\sqrt{2})$.  The estimation serves one of the tests of our approach.  The subject was studied in the same approach presented in \cite{yam} but the result is omitted in this paper.  We just point out that using the mass as the basic parameter makes the estimation of $\beta_{c}$ be just one of similar tasks, the estimation of the critical exponents and the amplitudes.  The analysis presented in this work is a numerical one heavily based upon the high temperature series up to large orders, and widely applicable in other models beyond almost solved square Ising model.

The present work is organized as follows:  In the next section, we briefly explain the series expansion approach to the susceptibility.   The $\delta$-expansion is introduced in this section and explain how it affects the behaviors of large mass expansion.  We shall show that by the $\delta$-expansion the scaling behavior is approximately observed in the large $M$ expansion.  In the next section, we estimate the critical exponent $\gamma$ via the ratio of derivatives ${\cal R}_{\ell}=\chi^{(\ell+1)}/\chi^{(\ell)}$ $(\ell=0,1,2,3,\cdots)$.  The estimation is carried out based upon the $\delta$-expanded large mass series.  Then, we conclude this paper.

\section{Series expansion and $\delta$ expansion}
\subsection{Series expansion at large and small $M$}
Conventional definition of the susceptibility is given by
\begin{equation}
\chi:=\sum_{all ~n}<s_{0}s_{n}>.
\end{equation}
The behavior of $\chi$ near the critical point was investigated in many literatures and has long history to the present day (see \cite{mc}).   A recent comprehensive study on the high temperature expansion was done in \cite{bou,chan} and we refer the works for review and main source of known results.  

To begin with our discussion, let us briefly review the critical behavior of the susceptibility $\chi$.  Expansion around the critical point is conveniently done with the variable  
\begin{equation}
\rho=(1/\sinh 2\beta -\sinh 2\beta)/2.
\end{equation}
At the critical temperature $\beta=\beta_{c}=\frac{1}{2}\log(\sqrt{2}+1)$, $\rho=0$, and at high temperature where $\beta<\beta_{c}$, $\rho>0$.   At high temperature $\beta$ near $\beta_{c}$, 
it has been settled that
\begin{equation}
\chi\sim const\times  \rho^{-7/4}+const+const \times \rho^{1/4}+O(\rho\log \rho).
\end{equation}
Since the variable $\rho$ can be expanded in more conventional one,
\begin{equation}
\tau=\beta_{c}-\beta,
\end{equation}
such that
\begin{equation}
\rho=2\sqrt{2}\tau+2\tau^2+\frac{16\sqrt{2}}{3}\tau^3+O(\tau^4),
\end{equation}
we find
\begin{equation}
\chi\sim const\times \tau^{-7/4}+const\times \tau^{-3/4}+const+const\times \tau^{1/4}+O(\tau\log\tau).
\end{equation}

Our approach uses the expansion of $\chi$ in terms of $1/M$.  So it is necessary to rewrite the above behavior in terms of $M$.  For the purpose we need series expansion of $\beta$ in $M$ at small enough $M$ and the steps are explained below:   The mass to be used in the investigation is not a priori known.  There may be some candidates, the so-called exponential mass, second moment mass and maybe others.  We have used in the previous work \cite{yam} the second one, since it is straightforwardly calculable in wide class of spin systems and field theories.  However, in the 2D Ising model on the square lattice, the computation has been carried out just up to $25$th order.  The order is not so high for obtaining conclusive results.   We therefore use the exponential mass, which is exactly obtained in \cite{mont},
\begin{equation}
\xi^{-1}=-\log\tanh\beta-2\beta.
\label{xi}
\end{equation}

Some comments would be in order.  We remind the reader that the above result is given in the large separation limit of the two point function where the spins are sited on a horizontal or vertical line.  Here we do not test the another candidate of the exponential mass which is defined by the diagonal correlation function where the spins are sited on a diagonal line.  
Further, we note that the counter part of $\xi$, the mass in momentum space, is given by $2(\cosh \xi^{-1}-1)$.  This quantity is in very close to the second moment mass defined by $4\chi/\mu$ which is more accessible to obtain in general models.  For example the first $8$ terms in the large mass expansions of $\beta$ agree with each other.   Relying upon the quantitative agreement of the two variable, instead of $4\chi/\mu$, we use basic parameter $M$ defined by
\begin{equation}
M:=2(\cosh \xi^{-1}-1).
\label{mass}
\end{equation}
Then, by using (\ref{xi}) and (\ref{mass}), we have
\begin{equation}
M=e^{-2\beta}\coth\beta+e^{2\beta}\tanh\beta-2.
\label{mbeta}
\end{equation}
Use of (\ref{mbeta}) allows us expansion of $\beta$ in $1/M$ to an arbitral large order.  
Around the critical temperature, we find
\begin{equation}
M=16\tau^2+16\sqrt{2}\tau^3+O(\tau^4).
\end{equation}
Thus, inversion gives
\begin{equation}
\tau\sim \frac{1}{4}M^{1/2}+O(M),
\end{equation}
and
\begin{equation}
\chi\sim const\times M^{-7/8}+const\times M^{-3/8}+const+const\times M^{1/8}+O(M^{1/2}\log M).
\label{chi_m}
\end{equation}
This is the goal which we aim recovering from large $M$ expansion of $\chi$.

Though the logarithmic term exists in (\ref{chi_m}) at order $M^{1/2}$, we neglect the presence.  In our trial to the susceptibility scaling, we just assume the power like behavior near the critical point (the constant term in (\ref{chi_m}) is interpreted as the term with zero power),
\begin{equation}
\chi=C_{1}x^{q_{1}}+C_{2}x^{q_{2}}+\cdots,\quad x:=\frac{1}{M}
\end{equation}
where $q_{1}>0$ and $q_{1}>q_{2}>\cdots$.  In ordinary term
\begin{equation}
q_{1}=\frac{\gamma}{2\nu}=\frac{7}{8},
\end{equation}
and $q_{2}=3/8$ and $q_{3}=0$.  

The derivative to the order $\ell$ is given by
\begin{equation}
\chi^{(\ell)}=C_{1}(q_{1})^{\ell}x^{q_{1}}+C_{2}(q_{2})^{\ell}x^{q_{2}}+\cdots
\end{equation}
and the ratio of derivatives behaves in the scaling region as
\begin{equation}
{\cal R}_{\ell}=q_{1}+C_{2}/C_{1}(q_{2}/q_{1})^{\ell}(q_{2}-q_{1})x^{q_{2}-q_{1}}+\cdots.
\label{key1}
\end{equation}
Thus, at least formally, we find for any $\ell$,
\begin{equation}
\lim_{x\to \infty}{\cal R}_{\ell}=q_{1}.
\label{key2}
\end{equation}
It is interesting to note that, since $q_{2}/q_{1}<1$, larger $\ell$ makes the correction smaller.  This leads us to expect that the convergence is faster for larger $\ell$.  Of course, large $\ell$ enhances higher order contribution which involves $(q_{i}/q_{1})^{\ell}$ ($i$ stands for some large integer).  So, actually there is a limitation of such suppression mechanism at some $\ell$.  We like to study on the point under the approach within large $M$ expansion and find a suitable value of $\ell$ for the estimation of the critical exponent $q_{1}$.

We now turn to the series expansion at large $M$,
\begin{equation}
\chi=1+\sum _{n=1}a_{n}x^n,\quad x=\frac{1}{M}.
\label{chi_m_2}
\end{equation}
The expansion of $\chi$ in $\beta$ has been carried out to extremely large orders up to few thousands \cite{bou}.  We here use first $100$ terms.  The substitution of $\beta$, which is given in series expansion in $1/M$ via (\ref{mbeta}), gives (\ref{chi_m_2}).   The Table IV in Appendix shows the coefficients of the first $40$ terms.

Then, for $\ell=1,2,3,\cdots$,
\begin{equation}
\chi^{(\ell)}=\sum _{n=1}n^{\ell}a_{n}x^n,\quad x=\frac{1}{M}.
\end{equation}
The ratio of derivatives has expansion:
\begin{eqnarray}
{\cal R}_{0}&=&a_{1}x+(2a_{2}-a_{1}^2)x^2+O(x^3),\\
{\cal R}_{\ell}&=&1+2^{\ell}\frac{a_{2}}{a_{1}}x+\frac{2\times 3^{\ell}a_{1}a_{3}-2^{2\ell}a_{2}^2}{a_{1}^2}x^2+O(x^3),\quad (\ell=1,2,3,\cdots).
\end{eqnarray}
It is interesting to see the convergence radius, approximately predicted by the ratio analysis of coefficients.   As clearly shown in Fig1, ${\cal R}_{\ell}$ exhibits behaviors quite different for $\ell=0,1,2$ and $3,4$.  The former cases show steady convergence to $-8$ but the later cases show no sign of convergence.
\begin{figure}[h]
\centering
\includegraphics[scale=0.95]{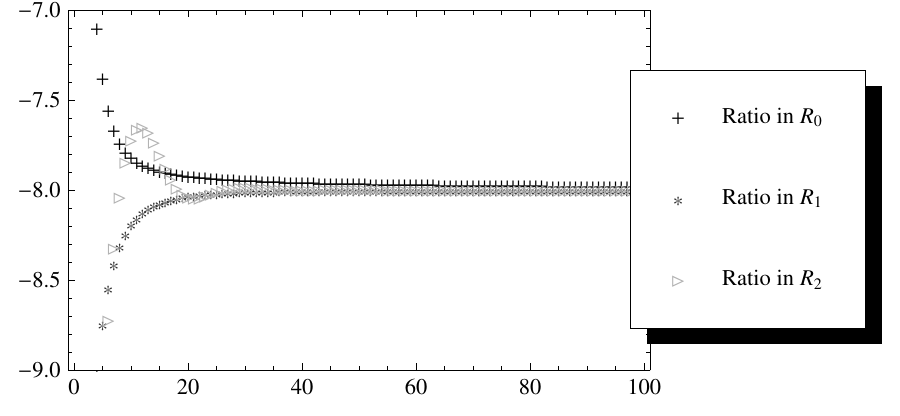}
\includegraphics[scale=0.95]{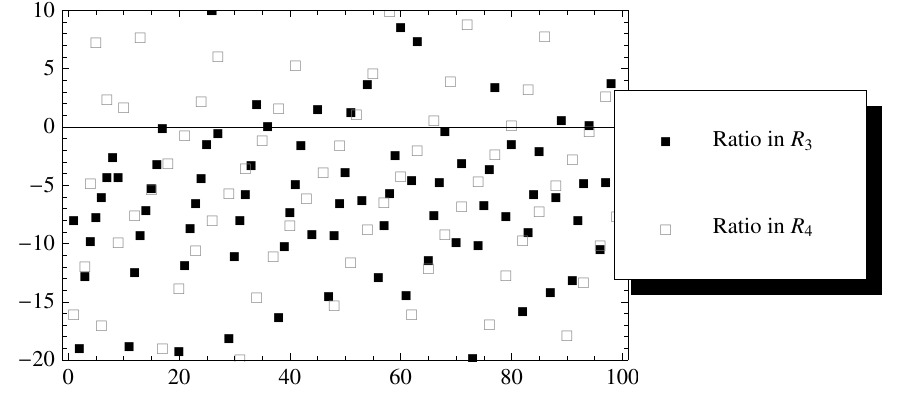}
\caption{Plot of ratio of the coefficients of series expansion of ${\cal R}_{\ell}=\chi^{(\ell+1)}/\chi^{(\ell)}$ ($\ell=0,1,2,3,4$).  The series of ${\cal R}_{\ell}$ for $\ell=0,1,2$ show alternative sign.  We have confirmed the alternative property up to $100$ terms.  All three serieses exhibit convergence to $-8$, which signals the radius of convergence $1/8=0.125$.  On the other hand, in ${\cal R}_{3}$ and ${\cal R}_{4}$, the coefficients do not show alternative property and the ratio fluctuate with large amplitude, thus exhibit no sign of convergence.  By these results, we expect that the series-property changes drastically at the jump from $\ell=2$ to $\ell=3$.   Note that these results are obtained solely from the high temperature expansion.}
\end{figure}

\subsection{Delta expansion}
Since the details of the $\delta$-expansion is discussed and explained in the past literatures, we do not repeat them.  Rather, we just state the essential idea and results and show the outline of the usage.  

The $\delta$-expansion is related to the dilation around the critical point $M=0$.  The dilation needs precise information of $\beta_{c}$ if $\tau$ is employed as the basic parameter.  However, in the study under the second order transition, the critical point is given in the mass as $M=0$ in the manner independent of the models.  Hence, the dilation is simply implemented when the basic parameter is chosen as the mass.  We make dilation in the thermodynamic function by the replacement of $M$ by $(1-\delta)/t$ where $t^{-1}$ stands for the rescaled mass parameter.  After the expansion of the function in $\delta$ up to a relevant order, setting $\delta=1$ gives a non-trivial dilated function in terms of $t$.  Due to the breaking of the regular correspondence between $M$ and $t$ at $\delta=1$, the physical interpretation of $t$ in this limit is obscured.  However, we are able to confirm that the limit $t\to \infty$ in the resultant function exactly recovers the correct limit of the original function (see for details \cite{yam,yam2}).  Denoting the $\delta$-expansion by the operation symbol $D$, the important result of $\delta$-expansion is summarized by
\begin{equation}
D[M^{-n}]=C_{N, n} t^n,\quad n=0,1,2,3,\cdots
\end{equation}
where $C_{N, n}$ denotes the binomial coefficient
\begin{equation}
C_{N, n}=\frac{N!}{n!(N-n)!}.
\end{equation}
Here we note that $N$ stands for the order of the full expansion.  In this sense, the $\delta$-expansion depends on the order of the large mass expansion.   If one faces with the comparison of the $\delta$-expanded $t$ series with the series valid near the critical point where $t\gg 1$, it is empirically known that the good matching occurs for the prescription
\begin{equation}
D[M^{p}]=C_{N, -p} t^{-p}=\frac{\Gamma(N+1)}{\Gamma(-p+1)\Gamma(N+p+1)}t^{-p}.
\end{equation}
From above we find for positive integer $p$ that $D[M^{p}]=0$.  This ensures the considerable suppression of the regular contribution involved in the expansion (\ref{chi_m}).

Thus, the $\delta$-expansion on the large mass expansion to $N$th order,
\begin{equation}
{\cal R}_{\ell}=a^{(\ell)}_{0}+\sum_{n=1}^{N}a^{(\ell)}_{n} x^n
\end{equation}
supplies the following series in $t$,
\begin{equation}
D[{\cal R}_{\ell}]=D\Big[a^{(\ell)}_{0}+\sum_{n=1}^{N}a^{(\ell)}_{n} x^n\Big]=a^{(\ell)}_{0}+\sum_{n=1}^{N}C_{N, n}a^{(\ell)}_{n} t^n:=\bar{\cal R}_{\ell}.
\end{equation}
Note that $a^{(\ell)}_{0}=0$ for $\ell=0$ and $a^{(\ell)}_{0}=1$ for $\ell\ge 1$.   As numerical check, we have drawn the behaviors of  ${\bar{\cal R}}_{\ell}$.   See Fig. 2, which shows the plots of ${\bar{\cal R}}_{\ell}$ at $\ell=0,1,2,3$.   Except for $\ell=3$, the $\delta$-expanded functions have effective regions roughly twice wider compared to those in the original series.   And we remark that the approach to the limit $q_{1}=7/8$ is convincing, which is the evidence that the $\delta$-expanded series contains within its effective region the scaling region (In the present case, the scaling region is the plateau with wide range).   
 \begin{figure}[h]
\centering
\includegraphics[scale=0.9]{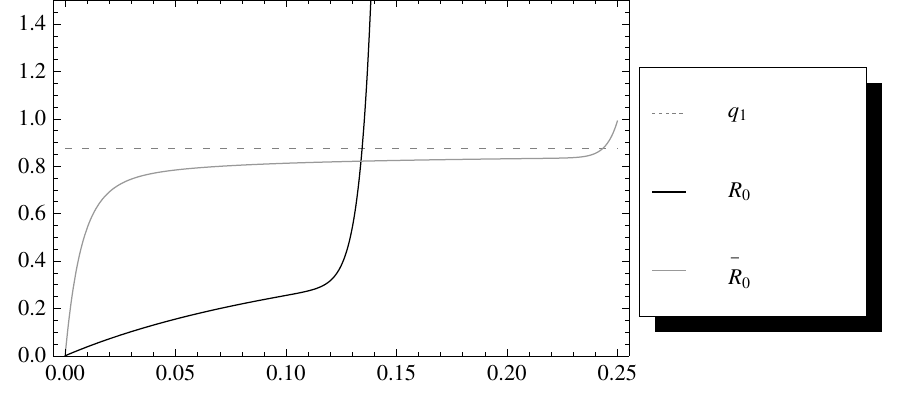}
\includegraphics[scale=0.9]{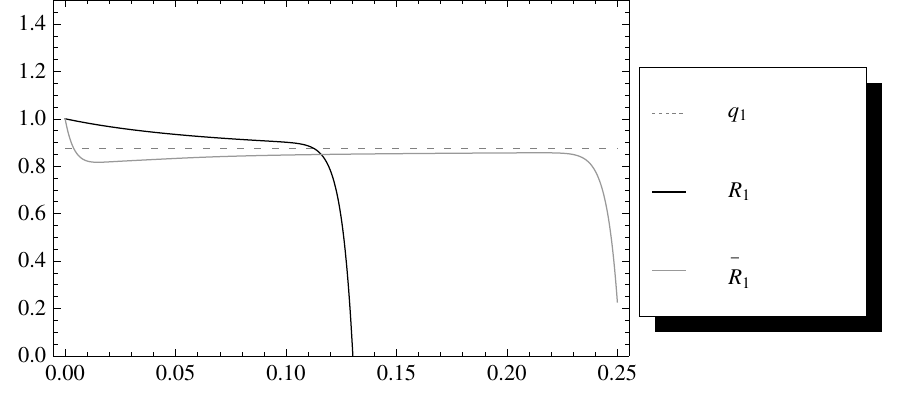}
\includegraphics[scale=0.9]{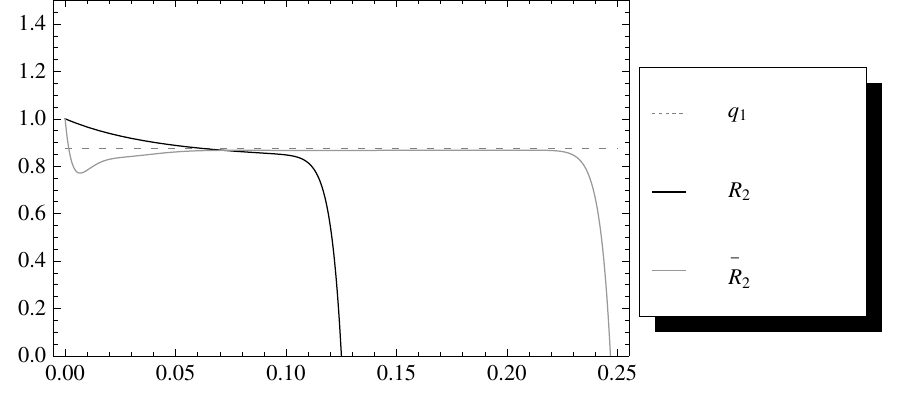}
\includegraphics[scale=0.9]{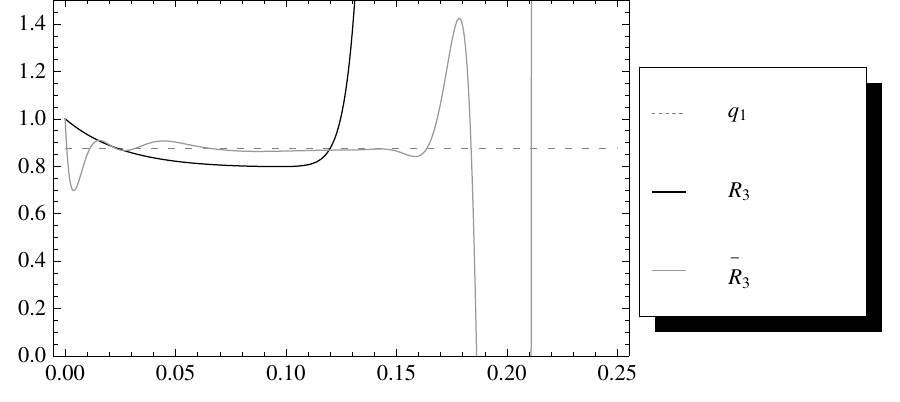}
\caption{The plot of ${\cal R}_{\ell}=\chi^{(\ell+1)}/\chi^{(\ell)}$ and $\bar{\cal R}_{\ell}=D[{\cal R}_{\ell}]$ $(\ell=0,1,2,3)$ at $25$th order in the large mass expansion.  The horizontal axis stands for $x$ for ${\cal R}_{\ell}$ and $t$ for ${\cal R}_{\ell}$.}
\end{figure}
It seems that $\bar{\cal R}_{2}$ gives best realization of $q_{1}$, though the reason why $\ell=2$ provides best behavior is not known to us.   The behavior of $\bar{\cal R}_{3}$ is oscillatory and not suitable for our purpose.  This is true also for $\bar{\cal R}_{4}$.  The first three plots share a common property that all of them has alternative coefficients (see also Tables V ~ IX in Appendix).

\section{Estimating $\gamma$ and sub-leading exponent}
\subsection{Non-parametric case}
In the previous section, we confirmed that the $\delta$-expansion revealed the behavior in the vicinity of the critical point $t=\infty$ at the region where $t$ is small.  This is the effect of dilation around the critical point.  Having prepared the estimation environment, let us set up the reference of the estimation by adopting naive use of the principle of minimum sensitivity (PMS) \cite{steve}.

First of all, we regard the plateau as the realization of the scaling region.  This identification is natural because it is quite conceivable that the asymptotic scaling behavior is just the convergence to the limit $q_{1}$ and stationary.   Actually, watching plot of $\bar{\cal R}_{0}$ (see Fig. 2), we see that the point of least variation would approximate the limit $q_{1}$.  Also for $\bar{\cal R}_{1}$ and $\bar{\cal R}_{2}$, the stationary points provide the approximation of $q_{1}$ due to the same reason.  In this manner, we can estimate $q_{1}$.  The result is summarized in Table I. 
\begin{table}[h]
\caption{Estimation result of $q_{1}=7/8=0.875$.}
\begin{center}
\begin{tabular}{ccccccccc}
\hline\noalign{\smallskip}
$\bar{\cal R}_{\ell}$  & $15$ & $20$ & $25$ & $30$ & $35$ & $40$ & $45$ & $50$\\
\noalign{\smallskip}\hline\noalign{\smallskip}
$\bar{\cal R}_{0}$   &  0.8169773  & 0.8276445   & 0.8317319  & 0.8370296 & 0.8391402 & 0.8424476 & 0.8437536 & 0.8460722  \\
$\bar{\cal R}_{1}$   & 0.8497474  & 0.8530335   &  0.8563755 & 0.8577449 & 0.8596025 & 0.8603685 & 0.8615928 & 0.8620879 \\
$\bar{\cal R}_{2}$   & 0.8661683  & 0.8662792   &  0.8675665 & 0.8679249 & 0.8686629 & 0.8689309 & 0.8694071 &0.8695930  \\
\noalign{\smallskip}\hline
\end{tabular}
\end{center}
\end{table}

As indicated in the plots in Fig. 2, the best result comes from $\bar{\cal R}_{2}$.  However, the accuracy is not satisfactory yet.  At $25$th order as a reference result, the error is about $1$ percent.  
For the accurate estimation of the exponent $q_{1}$, some additional device is needed to reduce the correction to the asymptotic scaling.  The device we employ here is the parametric extension of thermodynamic functions proposed in \cite{yam}.

\subsection{Parametric extension}
In the case of ${\cal R}_{\ell}$, the corresponding parametric extension gives
\begin{equation}
\psi_{\ell}(\alpha_{1}, \alpha_{2}, \cdots;x)=\Big\{1+\alpha_{1}x\frac{d}{dx}+\alpha_{2}\Big(x\frac{d}{dx}\Big)^2+\cdots\Big\}{\cal R}_{\ell}.
\end{equation}
We note that the differentiation deletes the leading constant $q_{1}$ (see (\ref{key1})).  Hence, 
irrespective of the values of $\alpha_{k}$, $\psi_{\ell}$ converges to $q_{1}=\gamma/2\nu$ as is easily understood from (\ref{key1}).   In the limit of $x\to \infty$, the independence of $\lim_{x\to \infty}\psi_{\ell}$ over $\alpha_{k}$ would be apparent, but in situation where the limit cannot be taken, the appropriate value of the parameters in the estimation work would exist.  Since $\psi_{\ell}\sim q_{1}+C_{2}/C_{1}(q_{2}/q_{1})^{\ell}(q_{2}-q_{1})(1+(q_{2}-q_{1})\alpha_{1}+(q_{2}-q_{1})^2\alpha_{2}+\cdots)x^{q_{2}-q_{1}}+\cdots$, the choice of parameters satisfying
\begin{equation}
1+(q_{2}-q_{1})\alpha_{1}+(q_{2}-q_{1})^2\alpha_{2}+\cdots=0
\label{cond}
\end{equation}
makes the leading correction vanishing.  If the $K$ parameters are introduced, it may be possible to delete or reduce considerably the first $K$ corrections.  In this work, however, we confine ourselves with the extension of just single parameter.   Then, it is apparent that the reduction of the leading correction needs the value of $q_{2}-q_{1}$, since (\ref{cond}) reads in this case
\begin{equation}
1+(q_{2}-q_{1})\alpha_{1}=0
\end{equation}
which yields $\alpha_{1}=-(q_{2}-q_{1})^{-1}$.  
Of course, we must work in the situation where the value of $q_{2}-q_{1}$ is not known to us.  Being blind on $q_{2}-q_{1}$, we must specify optimal value of $\alpha_{1}$ within the truncated large mass series.  For the task, it is crucial to consult with the $\delta$-expanded series where the scaling region is observed in the small $t$ expansion.   The effective reduction of the leading order correction would make plateau flatter than the parameter-less original function.  Thus, we extend the principle of minimum sensitivity (PMS) to fix optimal $\alpha_{1}$.  The step goes as follows:  First, we take that optimal value of $\alpha_{1}$ is given by the case where the stationary point of $\bar\psi_{\ell}$ becomes "maximally stationary".  The maximal stationarity means that, at the stationary point, following simultaneous conditions should hold,
\begin{eqnarray}
\bar\psi^{(1)}_{\ell}&=&\bar{\cal R}^{(1)}_{\ell}+\alpha_{1}\bar{\cal R}_{\ell}^{(2)}=0,
\label{firstcond}\\
\bar\psi^{(2)}_{\ell}&=&\bar{\cal R}^{(2)}_{\ell}+\alpha_{1}\bar{\cal R}_{\ell}^{(3)}=0.
\label{secondcond}
\end{eqnarray}
By imposing above conditions on the small $t$ series of $\bar\psi_{\ell}$, we may obtain optimal $\alpha_{1}=\alpha_{1}^*$ and the point $t=t^*$ at which $q_{1}$ should be estimated by $\bar\psi_{\ell}(\alpha_{1}^*,t^*)$.   However, in some cases, there exists no solution within plateau for the second condition (\ref{secondcond}), though the first condition (\ref{firstcond}) always has solution.  In this case, we instead take a loose condition which requires that the absolute value of the second derivative, $|\bar\psi^{(2)}_{\ell}|$, is minimum at the point where $\bar\psi_{\ell}$ is stationary.

Applying such a generalized PMS condition, we find sets of solutions $(\alpha_{1}^*, t^*)$.  In general the set is not unique at a given order.  Among them most reliable set would be the one with largest $t^*=t_{best}$.   The effectivity of this prescription manifests themselves by the fact that, at large orders, $t_{best}$ signals the limit at large-$t$ side of the plateau.  Thus, we obtain the best estimations by
\begin{eqnarray}
q_{1}&\sim& \bar\psi_{\ell}(\alpha_{1}^*, t_{best}),
\label{cond1}\\
(q_{1}-q_{2})^{-1}&\sim& \alpha_{1}^*.
\label{cond2}
\end{eqnarray}
The results are shown in Figs. 3, 4 and Tables II and III.
\begin{figure}[h]
\centering
\includegraphics[scale=1]{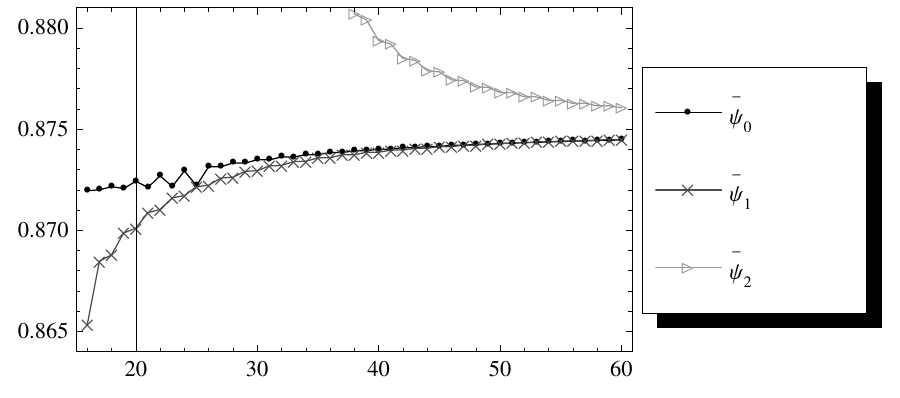}
\caption{The one-parameter estimation of $q_{1}=7/8$ with $\bar\psi_{\ell}$ ($\ell=0,1,2$).}
\end{figure}
\begin{figure}[h]
\centering
\includegraphics[scale=1]{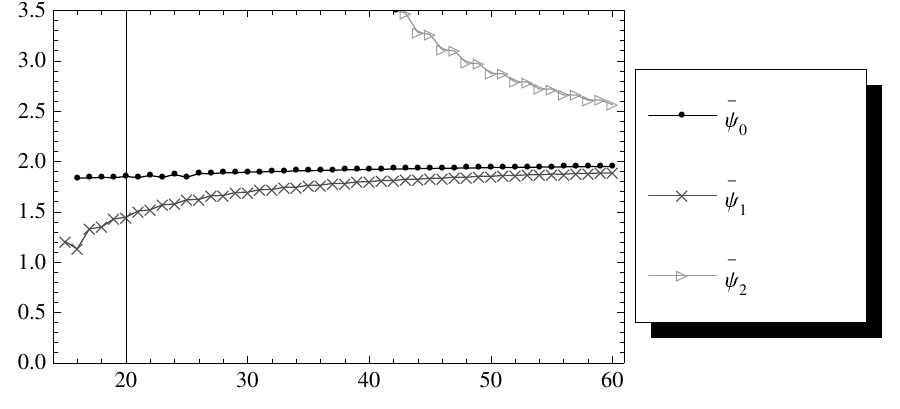}
\caption{The one--parameter estimation of $(q_{1}-q_{2})^{-1}=2$ with $\bar\psi_{\ell}$ ($\ell=0,1,2$).}
\end{figure}
\begin{table}[h]
\caption{Estimation result of $q_{1}=7/8=0.875$.  The case of $\ell=2$ has poor realization of the scaling behavior and estimation.  This is the reason of the blanks in the table. }
\begin{center}
\begin{tabular}{cccccccc}
\hline\noalign{\smallskip}
$\bar\psi_{\ell}$  & $20$ & $25$ & $30$ & $35$ & $40$ & $45$ & $50$ \\
\noalign{\smallskip}\hline\noalign{\smallskip}
$\bar\psi_{0}$   &  0.8724245  & 0.8722058   &0.8734845 & 0.8737348 & 0.8740136 & 0.8741421 & 0.8743013  \\
$\bar\psi_{1}$   & 0.8701054  & 0.8721568   &  0.8729444  & 0.8735773 & 0.8738560 & 0.8741360 & 0.8742653  \\
$\bar\psi_{2}$   &    &     &    &   & 0.8793704 & 0.8778297 & 0.8768158 \\
\noalign{\smallskip}\hline
\end{tabular}
\end{center}
\end{table}
\begin{table}[h]
\caption{Estimation result of $(q_{1}-q_{2})^{-1}=2$ via optimized $\alpha_{1}$.  The case of $\ell=2$ has poor realization of the scaling behavior and estimation.  This is the reason of the blanks in the table. }
\begin{center}
\begin{tabular}{cccccccc}
\hline\noalign{\smallskip}
$\bar\psi_{\ell}$  & $20$ & $25$ & $30$ & $35$ & $40$ & $45$ & $50$ \\
\noalign{\smallskip}\hline\noalign{\smallskip}
$\bar\psi_{0}$   &  1.8505000  & 1.8425165   & 1.8944572 & 1.9067215 & 1.9216758 & 1.9291058 &  1.9389608 \\
$\bar\psi_{1}$   & 1.4461756  & 1.6177268   &  1.6943348  & 1.7636010  & 1.7971446 & 1.8337286 & 1.8518406 \\
$\bar\psi_{2}$   &    &     &    &   & 3.8234502 & 3.2604537  & 2.8749102 \\
\noalign{\smallskip}\hline
\end{tabular}
\end{center}
\end{table}

By the numerical experiment, we find that, for $q_{1}$, the sequences exhibit clear convergence to the correct limit in $\bar \psi_{0}$ and $\bar\psi_{1}$.  For instance, $\bar\psi_{0}$ gives estimation $q_{1}\sim 0.8729$ (error$\sim 0.24$ percents) and $0.8722$ (error$\sim 0.32$ percents) at $N=24$ and $25$, respectively.   $\bar\psi_{2}$ implies slow convergence to the correct limit.  In the same manner with the $q_{1}$-sequence, the sequence of optimal $\alpha_{1}$ in $\bar\psi_{0}$ strongly indicates the correct value of $(q_{1}-q_{2})^{-1}=2$.  The cases $\ell=0,1$ provide similar results both of those are satisfactory.  On the other hand, the case $\ell=2$ ended in poor results.

\section{Concluding remarks}
To conclude our investigation, the $\delta$-expansion in the non-parametric scheme reveals that the cases $\ell=0,1,2$ manifest themselves that the scaling region emerges in the small $t$ region and the correct value of $q_{1}$ is indicated.  In the parametric cases with single parameter, the accuracy of $q_{1}$ and $(q_{1}-q_{2})^{-1}$ estimation is highly improved.  However, the case $\ell=2$ has failed in improving the accuracy.  Let us consider the reason of failure for the $\ell=2$ case.

The point is that, in the parametric scheme, the derivatives of ${\cal R}_{\ell}$ enters into the job.  For instance, we find from (\ref{cond1}) and (\ref{cond2}), the derivatives to the third one are needed to achieve the estimation procedures.  Then, for the success of the procedures, derivatives to the third one must show the approximate scaling around the estimation region of $t$ (plateau region of $\bar{\cal R}_{\ell}$).   So let us focus on the scaling behaviors of derivatives.  From (\ref{key1}), it follows that
\begin{equation}
\bar{\cal R}_{\ell}^{(k)}=(C_{2}/C_{1})C_{N,q_{2}-q_{1}}(q_{2}/q_{1})^{\ell}(q_{2}-q_{1})^k t^{q_{2}-q_{1}}+\cdots.
\end{equation}
To begin with we remind that for all $\ell$,
\begin{equation}
\lim_{t\to\infty}\bar{\cal R}_{\ell}^{(k)}=0.
\end{equation}
\begin{figure}
\centering
\includegraphics[scale=0.8]{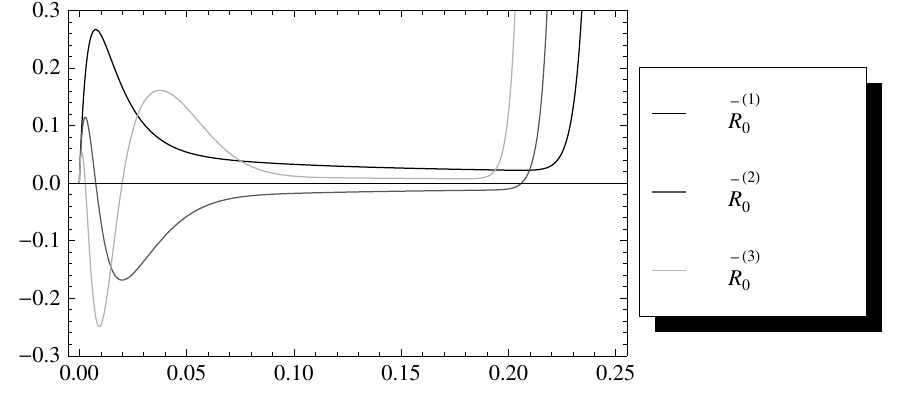}
\includegraphics[scale=0.8]{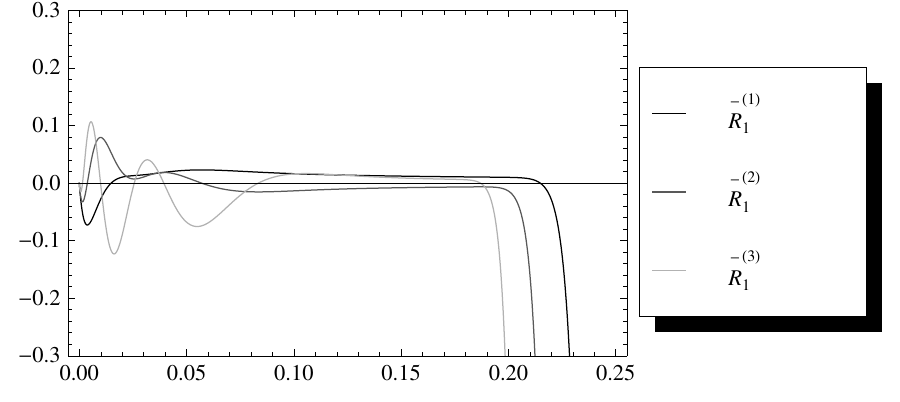}
\includegraphics[scale=0.8]{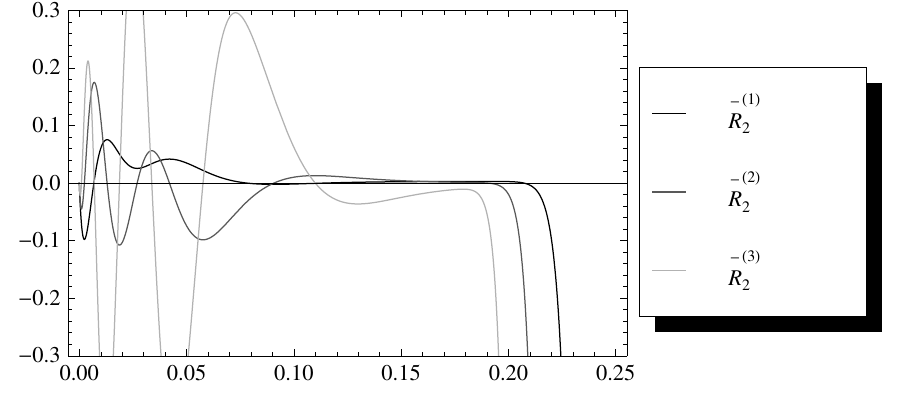}
\caption{$\bar{\cal R}_{\ell}^{(k)}$ for $\ell=0,1,2$ and $k=0,1,2,3$ at $25$th order.  For $\ell=0$ and $1$, the scaling region of derivatives are well developed.  On the contrary, for $\ell=2$, the scaling region is not developed yet.}
\end{figure}
Since $q_{2}-q_{1}<0$, the coefficient changes sign under the differentiation.  Then, both of $\bar{\cal R}_{0}^{(1)}$ and $\bar{\cal R}_{1}^{(1)}$ tend to zero from above and $q_{2}$ is found to be positive.   Hence, the sign of the front factor does not change with $\ell$.  As is understood from the plots of derivatives (see Fig. 5), in the cases $\ell=0,1$, those derivatives may be said as realizing the scaling.  However, in the case $\ell=2$, the derivatives do not show the scaling behavior yet.   In fact, scaling behavior of ${\cal R}_{2}^{(1)}$ is not clearly seen even at $100$th order (We note that these features can be observed solely from $\delta$-expanded small $t$ series.).  This is the reason why parametric extension of ${\cal R}_{2}$ does not bring about improvement.

As a final remark, we briefly compare our results in parametric extension with the traditional approaches.  Representative technique among them is the so-called Pad\'{e} approximant method of first derivative of $\log\chi$ in high temperature series ( series in $\beta$) \cite{peli}.     
The approach provides accurate $\beta_{c}$ (for instance $\beta_{c}\sim 0.4406838\cdots$ at $20$th order in the diagonal approximant) and, using the result, gives $\gamma\sim  1.7496\cdots$.  The accuracy is quite high.  Thus, our approach is not good in the accuracy of estimation.  There is one advantage in our approach, however.  In the traditional approaches such as the representative one has no unique estimation at a given order.  For example, there are other values of estimation depending on the choice of degrees of denominator and numerator of Pad\'{e} approximants.  On the other hand, we can identify which is the best one among a few candidates in our approach.  This selection becomes possible since the approximate critical behaviors of ${\cal R}_{\ell}$ becomes visible under the $\delta$-expansion.

There are related subjects not discussed in this work:  The parametric extension of the original thermodynamic quantities is not limited to the single parameter case.  Two- and three- parameter extension is a natural next step in our approach.  The study along this road is now under the progress.  We have found under yet rough examination  that the accuracy of estimation is further improved, but, unfortunately, the clear scaling begins to show at large order.  The method presented in this work is applicable also at low temperature.  It is also interesting to apply the present approach to other thermodynamic functions as the specific heat, the magnetization and the amplitude ratios and so on.  As another subject, the choice of the basic mass variable should also be studied.  Two candidate of the mass, second moment mass and the diagonal exponential mass, would be compared with each other on the results they would supply.   After the completion of these subjects, we like to report results in the subsequent publications.


\section*{Appendix}
\begin{table}[h]
\caption{Coefficients of $\chi(x)=1+\sum_{n=1} a_{n}x^n$ in the series expansion at $x\ll 1$ $(M\gg 1)$.}
\begin{center}
\begin{tabular}{cccc}
\hline\noalign{\smallskip}
$n$  & $a_{n}$  &$n$  & $a_{n}$ \\
\noalign{\smallskip}\hline\noalign{\smallskip}
$1$   &  4  & $21$   & 21275386763804768  \\
$2$   & -4  & $22$   &  -160006809343054864\\
$3$   & 16  & $23$   &  1206597595055984816\\
$4$   & -84  & $24$   &  -9121198369179912432\\
$5$   & 496  & $25$   &  69106863875292721536\\
$6$   & -3120  & $26$   &  -524680254679683977840\\
$7$   &  20416  & $27$   &  3991217801247779845008  \\
$8$   & -137300  & $28$   &  -30415396853284535164192\\
$9$   & 942368  & $29$   &  232169581608188044281504\\
$10$   & -6571808  & $30$   & -17749770132732227415243684\\
$11$   & 46422672  & $31$   &  13589773757322695502106928\\
$12$   & -331425504  & $32$   &  -104189603870648624732261332\\
$13$   & 2387361104  & $33$   &  799822083737000041347307488\\
$14$   & -17328288880  & $34$   &  -6147330757104867389693195232\\
$15$   &  126603329808  & $35$   &  47301411121280638308539291728  \\
$16$   & -930294191876  & $36$   &  -364358216053104960233960685968\\
$17$   & 6870391514160  & $37$   &  2809472823077732343284217635680\\
$18$   & -50965973697504  & $38$   &  -21684016505427268568571357447824\\
$19$   & 379584845946000  & $39$   &  167514287529323896647618797568880\\
$20$   & -2837208508428432  & $40$   &  -1295213666986423469602452039905120\\
\noalign{\smallskip}\hline
\end{tabular}
\end{center}
\end{table}

\begin{table}
\caption{Coefficients of $\chi^{(1)}/\chi^{(0)}=\sum_{n=1} a_{n}^{(0)}x^n$ in the series expansion at $x\ll 1$ $(M\gg 1)$.}
\begin{center}
\begin{tabular}{cccc}
\hline\noalign{\smallskip}
$n$  & $a_{n}^{(0)}$ & $n$  & $a_{n}^{(0)}$ \\
\noalign{\smallskip}\hline\noalign{\smallskip}
$1$   &  4  & $21$   & 1702823724672856736  \\
$2$   & -24  & $22$   &  -13501596184503027680 \\
$3$   & 160  & $23$   &  107091803029662561424 \\
$4$   & -1136  & $24$   &  -849705201298923243520 \\
$5$   & 8384  & $25$   &  6743862211146068408384 \\
$6$   & -63360  & $26$   &  -53538558608960941163360 \\
$7$   &  485888  & $27$   &  425140903442007642553648  \\
$8$   & -3760864  & $28$   &  -3376752124592333432399232 \\
$9$   & 29288176  & $29$   &  26826151806712841937551776 \\
$10$   & -229044704  & $30$   & -213159226806584799070943840 \\
$11$   & 1796618608  & $31$   &  1694067888051261417465988304 \\
$12$   & -14124299840  & $32$   &  -13465836756628212002914677632 \\
$13$   & 111232195728  & $33$   &  107055058630461662398852456816 \\
$14$   & -877185399072  & $34$   &  -851232593675722299370220140384 \\
$15$   &  6925285123760  & $35$   &  6769440346033398167021761676528  \\
$16$   & -54725053955264  & $36$   &  -53841525398447790365876496293504 \\
$17$   & 432786546475136  & $37$   &  428290955376655997494957586184224 \\
$18$   & -3424925637964512  & $38$   &  -3407332790404117881313638891492448 \\
$19$   & 27119362491604272  & $39$   &  27110758945124257909506817017689232 \\
$20$   & -214846102451691136  & $40$   &  -215733626693465249999762284727613824 \\
\noalign{\smallskip}\hline
\end{tabular}
\end{center}
\end{table}

\begin{table}
\caption{Coefficients of $\chi^{(2)}/\chi^{(1)}=1+\sum_{n=1} a_{n}^{(1)}x^n$ in the series expansion at $x\ll 1$ $(M\gg 1)$.}
\begin{center}
\begin{tabular}{cccc}
\hline\noalign{\smallskip}
$n$  & $a_{n}^{(1)}$ & $n$  & $a_{n}^{(1)}$  \\
\noalign{\smallskip}\hline\noalign{\smallskip}
$1$   &  -2  & $21$   & -5681180325529261272  \\
$2$   & 20  & $22$   &  45585039485774965592 \\
$3$   & -188  & $23$   &  -365622948610387580424 \\
$4$   & 1696  & $24$   &  2931554994773309328576 \\
$5$   & -14832  & $25$   &  -23498378896007252038632 \\
$6$   & 126800  & $26$   &  188309058357448190591800 \\
$7$   &  -1066536  & $27$   &  -1508735921272122489591320  \\
$8$   & 8866944  & $28$   &  12085838557513413548876288 \\
$9$   & -73099520  & $29$   &  -96799409028653454848531512 \\
$10$   & 598917240  & $30$   & 775193516919679155773697240 \\
$11$   & -4884337240  & $31$   &  -6207215526175544707533472648 \\
$12$   & 39692344624  & $32$   &  49698053761202350389581215232 \\
$13$   & -321670811568   & $33$   &  -397872043090052734523304056656 \\
$14$   & 2601181273480  & $34$   &  3185031697906826149624124991320 \\
$15$   &  -20997665014008  & $35$   &  -25494969986015979577666654717096  \\
$16$   & 169260623296960  & $36$   &  204065318897860700935135362721984  \\
$17$   & -1362814550375976  & $37$   &  -1633280963799008226225446117860920 \\
$18$   & 10962294305104952   & $38$   &  13071704883034769184711013037937368 \\
$19$   & -88109033749900600  & $39$   &  -104612959133323675229583441074445768 \\
$20$   & 707702384740998656  & $40$   &  837187318006362302642544817683192864 \\
\noalign{\smallskip}\hline
\end{tabular}
\end{center}
\end{table}

\begin{table}
\caption{Coefficients of $\chi^{(3)}/\chi^{(2)}=1+\sum_{n=1} a_{n}^{(2)}x^n$ in the series expansion at $x\ll 1$ $(M\gg 1)$.}
\begin{center}
\begin{tabular}{cccc}
\hline\noalign{\smallskip}
$n$  & $a_{n}^{(2)}$ & $n$ & $a^{(2)}$  \\
\noalign{\smallskip}\hline\noalign{\smallskip}
$1$   &  -4  & $21$   & -15040730135129492272  \\
$2$   & 56  & $22$   &  121004881555705996872 \\
$3$   & -640  & $23$   &  -972426936117955593200  \\
$4$   & 6480  & $24$   &  7803749063758302892096 \\
$5$   & -60224  & $25$   &  -62536717228676702935024 \\
$6$   & 525440  & $26$   &  500525348476468610236296 \\
$7$   &  -4373632  & $27$   &  -4002161071827691228836880  \\
$8$   & 35166208  & $28$   &  31979699003934529220811968 \\
$9$   & -275926528  & $29$   &  -255442783197624248573622832  \\
$10$   & 2131274216  & $30$   &  2040145044108510196576411400  \\
$11$   & -16329778128  & $31$   &  -16295163453848209150086894384 \\
$12$   & 124926689328  & $32$   &  130178881834711284732453163008 \\
$13$   & -959233922144  & $33$   &  -1040241930805335683349755457024 \\
$14$   & 7419013365176  & $34$   &  8314703351472082671211269216968 \\
$15$   &  -57904294231760   & $35$   &  -66476616064355872759999824896592  \\
$16$   & 456132757502656   & $36$   &  531596887846798342490795731022592  \\
$17$   & -3622486505331568  & $37$   &  -4251720114986581077363369713314544 \\
$18$   & 28950109717348424   & $38$   &  34008910282093517770619019429908936 \\
$19$   & -232333096023161872  & $39$   &  -272048440397204492233039080683024624 \\
$20$   & 1868720741877011520   & $40$   &  2176251364472264001414753196521168288 \\
\noalign{\smallskip}\hline
\end{tabular}
\end{center}
\end{table}

\begin{table}
\caption{Coefficients of $\chi^{(4)}/\chi^{(3)}=1+\sum_{n=1} a_{n}^{(3)}x^n$ in the series expansion at $x\ll 1$ $(M\gg 1)$.}
\begin{center}
\begin{tabular}{cccc}
\hline\noalign{\smallskip}
$n$  & $a_{n}^{(3)}$ & $n$  & $a_{n}^{(3)}$ \\
\noalign{\smallskip}\hline\noalign{\smallskip}
$1$   &  -8  & $21$   & -47163020009908433952  \\
$2$   & 152  & $22$   &  412116896023797485816 \\
$3$   & -1952  & $23$   &  -2699305167935251780896 \\
$4$   & 19216  & $24$   &  11968257756316458709056 \\
$5$   & -149568  & $25$   &  -17609179361962050886368 \\
$6$   & 904832  & $26$   &  -176308656907167515433512 \\
$7$   &  -3932160  & $27$   &  104349065828589559785376  \\
$8$   & 10355712  & $28$   &  3621304717086125252834028 \\
$9$   & -44546048  & $29$   &  -656440945752077901138957472  \\
$10$   & 1292483832  & $30$   & 7280425095959386195204357752  \\
$11$   & -24361161184  & $31$   &  -58726611083812091179172774560 \\
$12$   & 304417044592  & $32$   &  341020340137399083179694436352 \\
$13$   & -2844873583872  & $33$   &  -1119767807866941643624038227584 \\
$14$   & 20495746694344  & $34$   &  -2176067142302729239126317062056 \\
$15$   &  -109034249191392   & $35$   &  46891539299459592211807394564000  \\
$16$   & 349883316909760   & $36$   &  3292725713811400499298226387072  \\
$17$   & -50163076414560  & $37$   &  -6615630180814263178318159903485600 \\
$18$   & 1424993451633368   & $38$   &  108447822923488109588859786319705592 \\
$19$   & -206345171917851232  & $39$   &  -1109639966062292122355722655120070432 \\
$20$   & 3968211883960147136   & $40$   &  8153205060859798662287075542995817632 \\
\noalign{\smallskip}\hline
\end{tabular}
\end{center}
\end{table}

\begin{table}
\caption{Coefficients of $\chi^{(5)}/\chi^{(4)}=1+\sum_{n=1} a_{n}^{(4)}x^n$ in the series expansion at $x\ll 1$ $(M\gg 1)$.}
\begin{center}
\begin{tabular}{cccc}
\hline\noalign{\smallskip}
$n$  & $a_{n}^{(4)}$ & $n$  & $a_{n}^{(4)}$  \\
\noalign{\smallskip}\hline\noalign{\smallskip}
$1$   &  -16   & $21$   & -47163020009908433952  \\
$2$   & 392  & $22$   &  412116896023797485816 \\
$3$   & -4672  & $23$   &  -2699305167935251780896 \\
$4$   & 22224   & $24$   &  11968257756316458709056 \\
$5$   & 162304  & $25$   &  -17609179361962050886368 \\
$6$   & -2746240  & $26$   &  -176308656907167515433512 \\
$7$   &  -3932160  & $27$   &  104349065828589559785376  \\
$8$   & 10355712  & $28$   &  3621304717086125252834028 \\
$9$   & -44546048  & $29$   &  -656440945752077901138957472  \\
$10$   & 1292483832  & $30$   & 7280425095959386195204357752  \\
$11$   & -24361161184  & $31$   &  -58726611083812091179172774560 \\
$12$   & 304417044592  & $32$   &  341020340137399083179694436352 \\
$13$   & -2844873583872  & $33$   &  -1119767807866941643624038227584 \\
$14$   & 20495746694344  & $34$   &  -2176067142302729239126317062056 \\
$15$   &  -109034249191392   & $35$   &  46891539299459592211807394564000  \\
$16$   & 349883316909760   & $36$   &  3292725713811400499298226387072  \\
$17$   & -50163076414560  & $37$   &  -6615630180814263178318159903485600 \\
$18$   & 1424993451633368   & $38$   &  108447822923488109588859786319705592 \\
$19$   & -206345171917851232  & $39$   &  -1109639966062292122355722655120070432 \\
$20$   & 3968211883960147136   & $40$   &  8153205060859798662287075542995817632 \\
\noalign{\smallskip}\hline
\end{tabular}
\end{center}
\end{table}


\begin{thebibliography}{99}

\bibitem{yam} H. Yamada, ArXiv: 1303.3714 [hep-lat] (2013).
\bibitem{mc} B. M. McCoy, Advanced Statistical Mechanics, Oxford University Press, 2010.
\bibitem{peli} A. Pelissetto and E. Vicari, Phys. Rept. {\bf 368}, 549  (2002). 
\bibitem{bou} S. Boukraa, A. J. Guttmann, S. Hassani, I. Jensen, J.-M. Maillard, B. Nickel and N. Zenine, J. Phys. A {\bf 41}, 455202 (2008).  We have used the results of high temperature series opened to the public at http://www.ms.unimelb.edu.au/$\sim$iwan/ising/Ising\_ser.html.
\bibitem{chan} Y. Chan, A.J. Guttmann, B.G. Nickel and J.H.H. Perk, J. Stat. Phys. {\bf 145}, 549 (2011).
\bibitem{mont} E.W. Montroll, R.B. Potts and J.C. Ward, J. Math. Phys. {\bf 4}, 308 (1963).
\bibitem{yam2} H. Yamada, Phys. Rev. D {\bf 76} 045007 (2007).
\bibitem{steve} P. M. Stevenson, Phys. Rev. {\bf D23}, 2916 (1981).
\end{thebibliography}
\end{document}